\documentclass[%
 reprint,
 amsmath,amssymb,
 aps,
]{revtex4-2}

\usepackage{float}
\makeatletter
\let\newfloat\newfloat@ltx
\makeatother

\usepackage{graphicx}
\usepackage{dcolumn}
\usepackage{bm}
\usepackage{algorithm}
\floatname{algorithm}{NanoNets}

\usepackage{algpseudocode}
\usepackage{siunitx}

\begin{document}

\title{A kinetic Monte Carlo Approach for Boolean Logic Functionality \\ 
in Gold Nanoparticle Networks}

\author{Jonas Mensing¹}
\author{Wilfred G. van der Wiel²}
\author{Andreas Heuer¹}
\affiliation{
 ¹Institute of Physical Chemistry, University of Muenster, Germany \\
 ²NanoElectronics Group, MESA+ Institute for Nanotechnology, Center for Brain-Inspired Nano Systems (BRAINS), University of Twente, Enschede, The Netherlands,\\
 and Department of Physics, University of Muenster, Germany 
}

\begin{abstract}
\centering

Nanoparticles interconnected by insulating organic molecules exhibit nonlinear switching behavior at low temperatures. By assembling these switches into a network and manipulating charge transport dynamics through surrounding electrodes, the network can be reconfigurably functionalized to act as any Boolean logic gate. This work introduces a kinetic Monte Carlo-based simulation tool, applying established principles of single electronics to model charge transport dynamics in nanoparticle networks. We functionalize nanoparticle networks as Boolean logic gates and assess their quality using a fitness function. Based on the definition of fitness, we derive new metrics to quantify essential nonlinear properties of the network, including negative differential resistance and nonlinear separability. These nonlinear properties are crucial not only for functionalizing the network as Boolean logic gates but also when our networks are functionalized for brain-inspired computing applications in the future. We address fundamental questions about the dependence of fitness and nonlinear properties on system size, number of surrounding electrodes, and electrode positioning. We assert the overall benefit of having more electrodes, with proximity to the network's output being pivotal for functionality and nonlinearity. Additionally, we demonstrate a optimal system size and argue for breaking symmetry in electrode positioning to favor nonlinear properties. 

\end{abstract}

\maketitle


\section{\label{sec:intro}Introduction and Motivation\protect}

Brain-inspired computing represents an approach to advancing computation in machine learning applications by emulating the information processing of biological neural networks \cite{neuro_survey}. Compared to \textit{von-Neumann} computing, it leverages massively parallel operation without separating memory and processing \cite{neuro_algos}. The implementation of these brain-like infrastructures aim to overcome limitations of traditional hardware and ultimately reduce energy consumption of current machine learning applications \cite{Nakajima_2020, Markov_2014}.\\
Nanoparticle (NP) networks offer a promising avenue in this field. One current approach uses percolating scale-free NP networks with small-world properties \cite{brown1,brown2,brown3,brown4}. The intrinsic architecture of these networks includes conductivity dynamics which are crucial for computational tasks \cite{4298110, KAWAI201915}. Tunnel gaps forming between adjacent areas or clusters of nanoparticles act as nonlinear units or memristors, with their resistance exhibiting nonlinear responses to alterations in conductance. While the origin of the nonlinearity here arises from the connections of large clusters of NPs and not the particles themselves \cite{brown2}, our approach is different.\\ 
Our research in this domain centers single electronics within gold NP networks \cite{single_electronics_n_applications, DURRANI2003572}. Each NP serves as a conductive island, tunnel-coupled with its neighboring NP by an insulating organic molecule. When a NP is charged, its excess charges can tunnel to an adjacent NP only when the potential difference between them surpasses the repelling force exerted by the Coulomb energy. This \textit{Coulomb blockade} effect imbues the charge tunneling dynamics in between NPs with a nonlinear activation function \cite{double_quantum_dots}.\\
The network is surrounded by electrodes, which serve as network inputs, outputs or controls (see \mbox{FIG. \ref{fig:network_sketch}}). The latter manipulate the network's potential landscape and its charge tunneling dynamics, enabling the mapping of various functionalities to the input and output electrodes. Previous research has demonstrated the network's input to output dependence to reconfigurably function as any type of Boolean logic gate when applying suitable control electrode voltages \cite{bose2015, gold_nano2}. \mbox{FIG. \ref{fig:network_sketch}} shows a circuit diagram of a small NP  network and electrode setup.\\
The integration of theoretical frameworks and simulation methods is of key importance in order to enable a closer understanding and subsequent optimization of these systems. In this paper, we use a kinetic Monte Carlo (KMC) approach to simulate electronic charge tunneling dynamics within the NP network. Prior research has already demonstrated the effectiveness of the KMC approach for studying single electronics devices \cite{kmc1, kmc2, kmc3, kmc4}. In particular, Van Damme \textit{et al.} \cite{damme} functionalized small networks of $16$ nanoparticles to act as any Boolean logic gate. Our work extends previous analysis in several directions. First, it  addresses the influence of system size (number of NPs) as well as number and location of electrodes to optimize the chance to observe Boolean logic functionality. This information is relevant for the future design of such NP networks.  Second, from a methodological perspective  we introduce quantitative indicators for negative differential resistance and nonlinear separability. These nonlinear properties are imperative for a broad spectrum of brain-inspired computing \cite{Yi_2018} and machine learning applications \cite{1603620}. Here, we explicitly show that these indicators are quantitatively related to the ability of the network to display Boolean logic functionality. Third, due to the availability of fast array-based computations we highlight the efficiency of the implementation of the KMC approach.

\section{\label{sec:theory} Nanoparticle Networks}

In this section, we first describe some fundamentals of the NP network design of \cite{bose2015} which served as a starting point for our work on this simulation tool. We then go trough electrostatic properties of the network and describe the dynamics of single-electron tunneling.

\subsection{\label{sec:electrostatics}Electrostatics\protect}

\begin{figure}[t]
    \centering
    \includegraphics[width=0.45\textwidth]{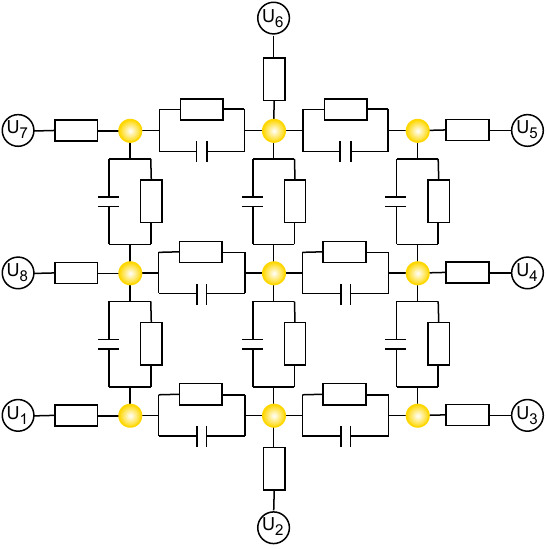}
    \caption{Circuit diagram of a gold nanoparticle network. Nanoparticles (golden dots) are tunnel-coupled with each other by organic insulators. The network represent a regular-grid like connection topology. A NP-NP connection is represented by a capacitor and resistor in parallel. The network is placed on a highly doped Si substrate with an insulating SiO$_{2}$ top-layer (not shown). The nanoparticles are surrounded by \mbox{$8$ Ti/Au electrodes} applying voltages $U_1$, \dots $U_8$.}
    \label{fig:network_sketch}
\end{figure}

The system of \cite{bose2015} consists of \mbox{Au NPs} which are interconnected by \mbox{1-octanethiols} as an insulating organic molecule. The network is placed on a highly doped Si substrate with an insulting \mbox{SiO$_{2}$ top-layer} and is surrounded by \mbox{$8$ Ti/Au electrodes}.\\
From an electrostatic point of view, we just speak of a network of conductive islands interconnected by resistors and capacitors in parallel. Electrodes are understood as voltage sources $U_E$. \mbox{FIG. \ref{fig:network_sketch}} shows a circuit diagram of a \mbox{$9$-NP} network in a regular grid-like connection topology.\\
We derive the capacitance values of the network using the image charge method (see \mbox{Supplementary Section \ref{sec:supp_i_c}}). The \mbox{capacitance $C_{ij}$} between \mbox{island $i$ and $j$} depends on the NP radius $r_{i}$, $r_{j}$, the distance between both NPs $d_{ij}$ and the permittivity of the insulating material $\epsilon_{\text{m}}$. The self capacitance $C_{i,\text{self}}$ for the isolated island $i$ close to the SiO$_{2}$ layer depends on its radius $r_{i}$ and the permittivity of the insulating environment $\epsilon_{\text{SiO}_{2}}$. Even though the simulation tool allows to setup variable radii or distances, for this work values are set constant with \mbox{$r = 10$ nm} and  \mbox{$d = 1$ nm}, respectively, following \cite{bose2015}.\\
For a given connection topology, we setup the capacitance matrix $\textbf{C}$ where diagonal components $(\textbf{C})_{ii}$ represent the sum of all capacitance values attached to \mbox{NP $i$} and its self capacitance, while off-diagonal components $(\textbf{C})_{ij}$ represent the cross capacitance in between \mbox{NP $i$ and $j$}. In this work, we will not cover the impact of having variable connection topologies. Therefore, for all results we align NP in a regular grid like two-dimensional topology, such as in \mbox{FIG. \ref{fig:network_sketch}}.

\subsection{\label{sec:se_tunneling}Single-Electron Tunneling\protect}

The network consists of conductive NPs interconnected by insulating molecules. The insulating molecules serve as tunnel barriers and functionalize the network. Excess charges residing on \mbox{NP $i$} tunnel to an adjacent \mbox{NP $j$} if their associated potential $\phi_i$ exceeds the repelling force resulting from the Coulomb energy $E_{C,j} = \frac{e^2}{(\textbf{C})_{jj}}$ of NP $j$. For a NP with 4 next neighbors at maximum we receive a charging energy of about \mbox{$E \approx 16$ meV}. This behavior leads to a switch-like nonlinear activation function for the charge tunneling dynamics where charges may or may not tunnel to their nearest neighbors corresponding to the current network's potential landscape.\\ 
We define the network state $\vec{q}(t)$ as the number of excess charges residing on each NP at a given time step.
Then, the network's potential landscape is calculated as
\begin{equation}
    \vec{\phi}(t) = \textbf{C}^{-1} \cdot \vec{q}(t)
\end{equation}
with $\textbf{C}^{-1}$ as the inverse of the capacitance matrix. The internal electrostatic energy is described as
\begin{equation}
    E = \frac{1}{2} \vec{q}(t) \cdot \vec{\phi}(t).
\end{equation}
When voltages are applied to the surrounding electrodes, charges either enter the network or getting drained from it. The Helmholtz's free energy determines the tunneling process and consists of the difference of total energy stored inside the network $E$ and work $W$ done by external electrodes
\begin{equation}
    F = E - W.
\end{equation}
In its current state, the network will enter its next state associated to a change in free energy, either by a charge tunneling event from \mbox{NP $i$ to $j$}
\begin{equation}
    \Delta F_{ij} = e\left(\phi_j - \phi_i\right)
                     + \frac{e^2}{2}\left[(\textbf{C}^{-1})_{ii}+(\textbf{C}^{-1})_{jj}-2(\textbf{C}^{-1})_{ij}\right]
    \label{eq:free_energy1}
\end{equation}
or by a charge tunneling event from NP $i$ to electrode $E$ 
\begin{equation}
    \Delta F_{iE} = e\left(U_e - \phi_i\right) + \frac{e^2}{2}(\textbf{C}^{-1})_{ii},
    \label{eq:free_energy2}
\end{equation}
with elementary charge $e$.
The change in free energy caused by a tunnel event serves as a measure of probability for this specific event with tunnel rate
\begin{equation}
    \Gamma_{ij} = - \frac{\Delta F_{ij}}{e^2 R_{ij}}\cdot\left[1 - \exp\left(\frac{-\Delta F_{ij}}{k_B T}\right)\right]^{-1},
    \label{eq:tunnel_rate}
\end{equation}
where $R_{ij}$ denotes the resistance for jump path $i \rightarrow j$ and $T$ the network's temperature. 
If we apply constant voltages to all electrodes and execute at maximum about $10,000$ charge tunneling events, the network will eventually settle in an equilibrium with constant electric currents entering or exiting the system via the electrodes.\\
We revisit \mbox{Eq. (\ref{eq:tunnel_rate})} and see that increasing the network's temperature $T$ leads to drastically larger rates. Then, differences in the potential landscape are negligible and thermal excitation will be the dominant factor for the charge tunneling through the network. Overall, the device looses its functionality due to the linearization of the charge tunneling dynamics, i.e. its nonlinear activation functions (see Supplementary Section \ref{sec:supp_iv} and \mbox{FIG. \ref{fig:iv_curves}}). Therefore, we have to choose the temperature, so that condition $E_C > k_B T$ is true. In this work we will stick to a temperature value of $T = 0.28$ K as in \cite{bose2015} which easily satisfies the latter condition with \mbox{$k_B T \approx 2.5 \cdot 10^{-2}$ meV} much smaller than the maximum charging energy of \mbox{$E \approx 16$ meV}.\\
Additionally we also want to neglect coherent quantum processes, i.e. co-tunneling. For this we have to choose the tunneling resistance $R_{ij}$ in between NPs to be much higher than the quantum resistance of \mbox{$R_t = \frac{h}{e^2} \approx 25.8 \text{ k$\Omega$}$}. In our simulations we set the tunneling resistance for all tunneling processes to a constant value of \mbox{$R = 25 \text{ M$\Omega$}$}, which is sufficient to assume charges to be confined on our NP islands.

\section{\label{sec:methods}Modelling and Simulation\protect}

Based on the findings in the previous section, we explain how to build a KMC simulation tool to model the charge tunneling dynamics in NP networks. We also explain how to functionalize the network to mimic Boolean logic gates. In the last part we define the fitness value as a way to evaluate the quality of the network behaving a a particular Boolean logic gate. Based on this quality factor we derive $Q_{\text{NDR}}$ and $Q_{\text{NLS}}$ as new quantities to measure the properties of negative differential resistance (NDR) and nonlinear separability (NLS). 

\begin{algorithm}[t]
    \caption{Simulation Tool}\label{alg:simulation_tool}
    
    \begin{algorithmic} [1]
        \State Initialize network topology
        \State Attach electrodes and define $U_E$
        \State Initialize capacitance matrix $\textbf{C}$
        \While{$u_I \ge u_{th}$}
            \State Compute $\Delta F_{ij}(\vec{\phi}(t))$
            \State Compute $\Gamma_{ij}(\Delta F_{ij})$
            \State Compute $CDF_{\Gamma}(n)$ and select $n$-th event
            \State $t \rightarrow t+1$
            \State $\vec{\phi}(t) \rightarrow \vec{\phi}(t+1)$
        \EndWhile
    \end{algorithmic}
    
\end{algorithm}

\subsection{\label{sec:kmc}Kinetic Monte Carlo Method\protect}

For a given network topology of $N_{\text{NP}}$ nanoparticles, initialized capacitance matrix $\textbf{C}$ and set of electrode voltages $U_E$, we firstly initialize the network's state \mbox{$\vec{q} = \vec{0}$} and potential landscape \mbox{$\vec{\phi} = \vec{0}$}. Then for all possible tunneling events we calculate the change in free energy by \mbox{Eq. (\ref{eq:free_energy1}) and (\ref{eq:free_energy2})} and tunnel rates by \mbox{Eq (\ref{eq:tunnel_rate})}.\\
In the next step we compute the cumulative distribution function (CDF) of all tunnel rates $CDF_\Gamma(n)$. The $CDF$ consists of $N_T$ steps, one for each charge tunneling event $n$ with its last step $CDF_\Gamma(n=N_\text{T})$ as the total rate constant $k_\text{tot}$. Now, for the KMC procedure we first need to sample two random numbers $r_1$ and $r_2$ and find $n$ where

\begin{equation}
    CDF_\Gamma(n-1) < r_1 \cdot k_{\text{tot}} \le CDF_\Gamma(n).
    \label{eq:select_event}
\end{equation}

As $n$ was selected, we execute its corresponding charge tunneling event from \mbox{NP $i$} to \mbox{NP $j$}. This will leave the current state $\vec{q}(t_1)$ towards $\vec{q}(t_2)$. Afterwards we update the time scale

\begin{equation}
    t_2 = t_1 - \ln\left(\frac{r_2}{k_{\text{tot}}}\right).
    \label{eq:time}
\end{equation}

\begin{figure}
\centering
\includegraphics{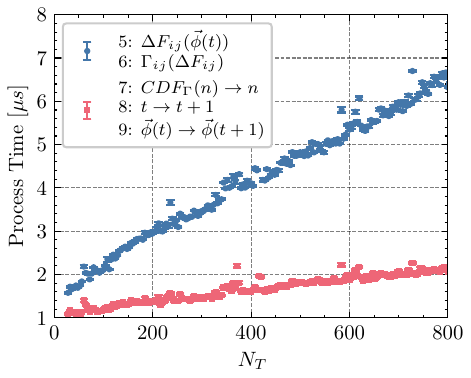}
\caption{\label{fig:complexity}
Process time for calculating all tunnel events (blue curve) and advancing into the next step by selecting an event (red curve).}
\end{figure}

Since only two elements of the state vector have been updated, for the next potential landscape we can just evaluate

\begin{equation}
    \phi_{t_2} = \phi_{t_1} + \textbf{C}^{-1} \cdot \vec{j}
    \label{eq:next_pots}
\end{equation}

where $\vec{j}$ is a vector including zeros and elementary charge $-e$ at \mbox{position $i$} and $+e$ at \mbox{position $j$}.\\
\textbf{NanoNets} shows an overview of the KMC procedure. We implemented the simulation tool in Python using \textit{NumPy} \cite{numpy} and \textit{Numba} \cite{numba} for fast array-based computations (for \textbf{NanoNets} GitHub see \cite{NanoNets}). The process of updating potentials and selecting the next event from the sorted $CDF$ is achieved in about $O(\log(N))$. However, since the whole system is coupled via the potential landscape, we have to calculate all tunnel rates after each tunneling event, thus changed potential landscape. This leads to a time complexity of about $O(N)$. \mbox{FIG. \ref{fig:complexity}} shows the process time for the major parts in the KMC procedure.

\subsection{\label{sec:comp_funcs}Computing Functionalities\protect}

In this paper we want to investigate computing functionalities in the form of Boolean logic gates as a systematic property of the NP network. For Boolean logic we need two inputs and one output. Inputs, either set on/1 or off/0, produce an output signal depending on the underlying logic operation.\\
In our case, two electrodes will always serve as input electrodes. These electrodes are set to \mbox{$U_I \in \{0.0, 10.0\} \text{ mV}$}. Accordingly, there is an output electrode, which is always grounded, thus \mbox{$U_O = 0.0 \text{ V}$}. At the output we will measure the electric current

\begin{equation}
    I_O = \frac{e \cdot (N_{\text{Net} \rightarrow O} - N_{O \rightarrow \text{Net}})}{t} 
\end{equation}

as a function of time $t$ passed in the KMC procedure and number of elementary charge jumps from network to output $N_{\text{Net} \rightarrow O}$ and from output to network $N_{O \rightarrow \text{Net}}$. All other electrodes are called control electrodes with voltages set inside a range of \mbox{$U_C \in [-50.0,50.0] \text{ mV}$}. 
The system's phase space is spanned by \mbox{$N_C$} electrodes with each point in phase representing a set of four different electric currents $(I_{00},I_{10},I_{01},I_{11})$ corresponding to the four possible input electrode voltage combinations.\\
When increasing the network size, the electrode voltages drop across an increasing amount of NPs. To ensure comparable results, we scale the voltages of control and input electrodes to prevent the electric current at the output electrode from simply converging to zero as the network expands. To achieve this, we conducted experiments to assess the dependencies between input electrode voltage and output electrode electric current (I-V curves) across multiple systems. Through this analysis, we derived a scaling factor that maintains the electric current at a consistent level when multiplied by the initial voltage value, see \mbox{Supplementary Section \ref{sec:supp_scaling}} and \mbox{FIG. \ref{fig:volt_scaling}}.\\
Summarizing the simulation procedure, we first initialize a network of regular grid-like topology, electrode positions and electrostatics. Then we apply constant voltages to all electrodes and evolve the system into its equilibrium state. Afterwards, we start to track the time scale and the number of charge jumps exiting and entering the system via the grounded output electrode. This allows us to compute the electric current $I$ and its relative uncertainty $u_{I}$. The simulation ends when the KMC procedure reaches $u_{I} = 0.05 = 5 \%$ uncertainty or when ten million charge tunneling events have been executed.

\subsection{\label{sec:ana_methods}Analysis Methods\protect}

For a single simulation run, we sample $20,000$ different control 
combinations, resulting in a set of four electric currents $(I_{00},I_{10},I_{01},I_{11})$ for each combination, totaling $80,000$ currents. To characterize nonlinear properties we introduce the three  parameters

\begin{eqnarray}
\label{eq:vardef}
M_{\rm l} & = & \frac{1}{4} (I_{11} - I_{01} + I_{10} - I_{00} ) \nonumber\\
M_{\rm r} & = & \frac{1}{4} (I_{11} + I_{01} - I_{10} - I_{00} ) \\
X & = & \frac{1}{4} (I_{11} - I_{01} - I_{10} + I_{00} ).\nonumber
\end{eqnarray}

$M_l$/$M_r$ signify the increase in output current upon altering the first/second input voltage, whereas $X$ quantifies the cross-correlation between the two input voltages. In qualitative terms, $M_l$ and $M_r$ can be interpreted as an effective mobility of the output current concerning one of the two input voltages, while $X$ denotes a measure of nonlinear coupling between both inputs.\\
Our objective is to evaluate the network's ability to be reconfigured into each Boolean logic gate using the fitness function

\begin{equation}
    F = \frac{m}{\sqrt{MSE} + \delta \cdot |c|}    
    \label{eq:fitness}
\end{equation}

as a quality metric for the accuracy with which a given set of currents $(I_{00},I_{10},I_{01},I_{11})$ represents a specific gate \cite{bose2015}. The fitness comprises the slope or signal $m$, the mean squared error or noise $MSE$, and the absolute offset $|c|$. The signal is defined as \mbox{$m = I_{on} - I_{off} = \frac{\sum_{i \in ON} I_i}{|ON|} - \frac{\sum_{j \in OFF} I_j}{|OFF|}$} where $ON$/$OFF$ represents the set of electric currents corresponding to the gate's on/off currents, and \mbox{$|ON|$/$|OFF|$} is the cardinality. For example, for an AND gate, \mbox{$ON = \{I_{11}\}$} and \mbox{$OFF = \{I_{00},I_{10},I_{01}\}$}. The noise is defined as \mbox{$MSE = \frac{1}{4}[\sum_{i \in ON} (I_i - I_{on})^2 + \sum_{j \in OFF} (I_j - I_{off})^2]$} and the offset as \mbox{$c = I_{off}$}. In total, the fitness consists of a \textit{signal-to-noise} ratio where logic gates with small offsets are more favourable from the experimental perspective as expressed by $\delta > 0$. Theoretically it is easier to choose $\delta = 0$. In \mbox{Supplementary Section \ref{sec:supp_delta}} and \mbox{FIG. \ref{fig:delta_dep}} we cover the influence of $\delta > 0$.\\
For $\delta = 0$ we can exactly rewrite \mbox{Eq. (\ref{eq:fitness})} for the individual gates in form of the quantities introduced in \mbox{Eq. (\ref{eq:vardef})}

\begin{widetext}
\begin{eqnarray}
\label{eq:fitness_new}
F_{\text{AND}} = &-F_{\text{NAND}} &=  \sqrt{\frac{8}{3}} \frac{M_l + M_r + X}{\sqrt{M_l^2 + M_r^2 + X^2 - M_lM_r - M_lX - M_rX}} \nonumber\\
F_{\text{OR}} = &-F_{\text{NOR}}  &=  \sqrt{\frac{8}{3}} \frac{M_l + M_r - X}{\sqrt{M_l^2 + M_r^2 + X^2 - M_lM_r + M_lX + M_rX}} \nonumber\\
F_{\text{XOR}} = &-F_{\text{XNOR}} & = \frac{-2X}{\sqrt{M_l^2 + M_r^2}}
\end{eqnarray}
\end{widetext}

Our goal is to estimate the first and second moment of the fitness distribution for each gate in terms of the statistical properties of $M_l$, $M_r$, and $X$. For this purpose we use two approximations:  Firstly, we approximate $<F(x)>$ by  $F(<x>)$ and, secondly, we neglect  $\text{cov}(M_l,X)$ and $\text{cov}(M_r,X)$. Both terms are negligible relative to $\text{cov}(M_l,M_r)$, as demonstrated in Supplementary Section \ref{sec:supp_paras} and \mbox{FIG.\ref{fig:x_and_r_dist}}.  Furthermore, since we only consider regular grid-like networks with a symmetric electrode connection, we naturally have $\langle X \rangle = 0$ as well as identical distributions for $M_l$ and $M_r$ when averaging over a sufficiently larger number of control voltages.
Therefore, we may abbreviate \mbox{$<M_l> = <M_r> = <M>$} and \mbox{$<M_l^2> = <M_r^2> = <M^2>$}.\\
Using these approximations, for the AND, OR, NAND, and NOR gates we can  write 

\begin{widetext}
    \begin{eqnarray}
        \label{eq:f_expected}
        <F_{\text{AND/OR}}> \propto & \frac{<M>}{\sqrt{<X^2> + <M^2> + \text{Var}(M)\cdot(1 - corr(M_l,M_r))}} & \propto -<F_{\text{NAND/NOR}}>\nonumber\\
        <F_{\text{AND/OR}}^2> \propto & \frac{<X^2> + <M^2> + \text{Var}(M)\cdot(1 + corr(M_l,M_r))}{<X^2> + <M^2> + \text{Var}(M)\cdot(1 - corr(M_l,M_r))} & \propto <F_{\text{NAND/NOR}}^2>\nonumber\\
        \text{Var}(F_{\text{AND/OR}}) \propto & \frac{<X^2> + \text{Var}(M)\cdot(2 + corr(M_l,M_r))}{<X^2> + <M^2> + \text{Var}(M)\cdot(1 - corr(M_l,M_r))} & \propto \text{Var}(F_{\text{NAND/NOR}})
    \end{eqnarray}
\end{widetext}

and for the XOR and XNOR gates

\begin{eqnarray}
    <F_{\text{XOR/XNOR}}> &=&  0 \nonumber\\
    <F_{\text{XOR/XNOR}}^2> &\propto&  \frac{<X^2>}{<M^2>} .\nonumber\\
    \label{eq:f_expected2}
\end{eqnarray}

\begin{figure*}
\centering
\includegraphics[width=\textwidth]{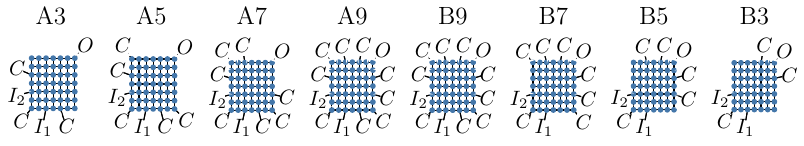}
\textbf{a)}
\includegraphics[width=0.65\textwidth]{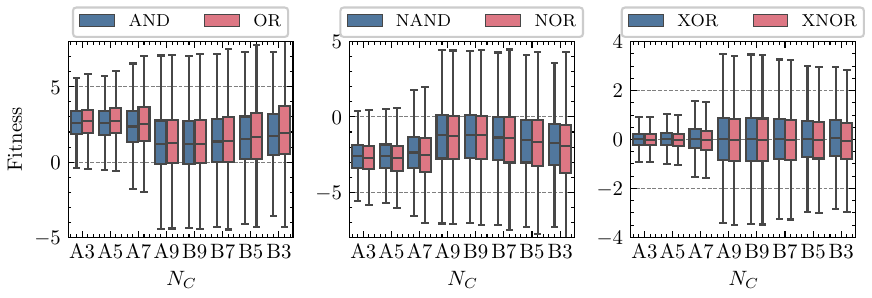}
\textbf{b)}
\includegraphics[width=0.29\textwidth]{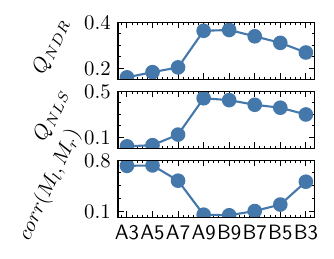}
\caption{\label{fig:controls_impact}
\textbf{a)} Box plots of fitness values for all logic gates for networks with varying numbers of control electrodes. The length of a bar is equal to two standard deviations. Each box corresponds to one of the network diagrams shown above. The networks are projected along the x-axis in first ascending and then descending order of control electrode numbers. \textbf{b)} Measure of negative differential resistance ($Q_{\text{NDR}}$), nonlinear separability ($Q_{\text{NLS}}$), and Pearson correlation $corr(M_l,M_r)$ across variable numbers of control electrodes.}
\end{figure*}

Nonlinear behavior is in particular relevant for the NAND, NOR, XOR, and XNOR gates. For these gates we want to define measures which reflect the likelihood to find gates with sufficiently high fitness. For the NAND/NOR gates we consider the ratio \mbox{$<F_{\text{NAND/NOR}}>/\sqrt{\text{Var}(F_{\text{NAND/NOR}})}$}. Both, if \mbox{$<F_{\text{NAND/NOR}}>$} is not too  negative and/or the standard deviation \mbox{$\sqrt{\text{Var}(F_{\text{NAND/NOR}})}$} is sufficiently large, there is a higher probability that positive fitness values occur, i.e. the system starts to display the respective functionality. On this basis we define

\begin{widetext}
    \begin{eqnarray}
        Q_{\text{NDR}} &=& \frac{1}{2} \cdot \left[ 1 - tanh\left(\frac{<F_{\text{AND/NAND}}>}{\sqrt{\text{Var}(F_{\text{AND/NAND}})/2}}\right) \right] \nonumber\\
                &=& \frac{1}{2} \cdot \left[1 - tanh\left(\frac{<M>}{\sqrt{<X^2>/2 + \text{Var}(M)\cdot(1 + corr(M_l,M_r)/2)}}\right)\right]\nonumber\\
                &\stackrel{\text{Var}(M) \gg <X^2>/2}{\approx}& \frac{1}{2} \cdot \left[1 - tanh\left(\frac{<M>}{\sigma(M)}\right)\right].
        \label{eq:ql}
    \end{eqnarray}
\end{widetext}

In the last step we have neglected the Pearson correlation between $M_l$ and $M_r$.\\
The properties of the $\tanh$-function mathematically restrict  $Q_{\text{NDR}}$ to values between \mbox{$0$ and $1$}. Since $\langle M \rangle$ is always positive,  the actual upper bound is 0.5. For values of $Q_{\text{NDR}}$ close to zero, the possibility of NAND and NOR gates is very small. Qualitatively, this measure allows us to assess the nonlinear feature of \textit{negative differential resistance} (NDR) required to achieve NAND/NOR gates.\\
We mention in passing that the variance in \mbox{Eq. (\ref{eq:f_expected})} becomes larger for stronger correlations between $M_l$ and $M_r$. This increases the probability to find gates (AND, OR, NAND, NOR) with  higher fitness values.\\
From \mbox{Eq. (\ref{eq:f_expected2})} we may conclude that XOR and XNOR gates have the same statistical properties. Since in this case the likelihood to find well-performing exclusive gates with high fitness values is directly related to the size of the variance of the fitness distribution, we can introduce

\begin{equation}
    Q_{\text{NLS}}  = \frac{<X^2>}{<M^2>}
    \label{eq:qlr}
\end{equation}

as a general measure for \textit{nonlinear separability} (NLS), strongly reflecting the size of the nonlinear coupling $X$ between both inputs.

\section{\label{sec:results}Results\protect}

In our study of nanoparticle networks manipulated by surrounding electrodes, key design features include the number of nanoparticles ($N_{\text{NP}}$) and the location and quantity of control electrodes ($N_{C}$). In this section, we first analyze the dependence of logic gate fitness ($F$) on the location and number of control electrodes. Given the significant impact of electrode positioning, the subsequent section addresses an increase in system size while considering two distinct electrode positioning setups. All results are contextualized within the framework of nonlinear properties. Additional insights into parameter dependence on $N_{\text{NP}}$ and $N_{C}$ can be found in \mbox{Supplementary Section \ref{sec:supp_paras}} and \mbox{FIG. \ref{fig:fitness_parameter}}.

\subsection{Number and Positioning of Control Electrodes}
\label{sec:controls}

\begin{figure*}
\centering
\includegraphics[width=\textwidth]{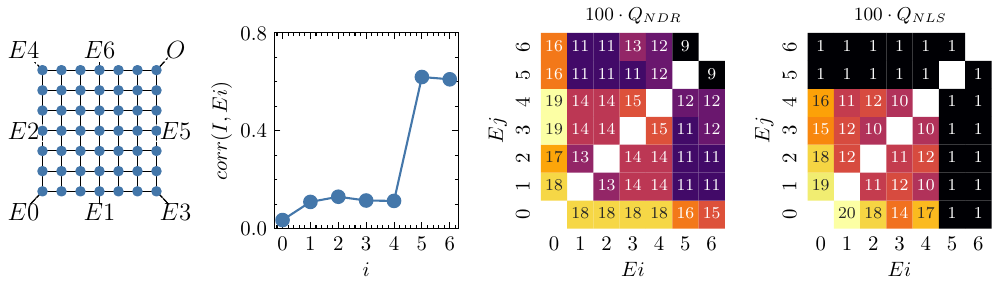}
\caption{\label{fig:electrode_sens}
Influence of input electrode positions. The graph depicts the simulated network. For every electrode combination $(E_i,E_j)$, we examine four scenarios $(E_i,E_j),(E_i+\Delta,E_j),(E_i,E_j+\Delta),(E_i+\Delta,E_j+\Delta)$. The figures show current-voltage correlation, negative differential resistance (NDR), and nonlinear separability (NLS). To display annotations as integers in the latter two, each cell is scaled by a factor of $100$.}
\end{figure*}

In this section, we explore the impact of the number of control electrodes and their positioning on network functionality, focusing on regular grid-like networks as discussed in \mbox{Section \ref{sec:electrostatics}}. We use a $7 \times 7$ grid of NPs ($N_{\text{NP}} = 49$) with one output and two input electrodes. To ensure that the majority of the system resides between input and output, we connect the output to a corner NP, and both inputs to the opposite edges.\\
We start by adding control electrodes near both inputs as we move closer to the output. These simulations are denoted $AN_{C}$, where $N_C$ represents the number of controls. Subsequently, controls are removed from the system, starting with those near both inputs. These simulations are denoted $BN_{C}$. The upper sketches in \mbox{FIG. \ref{fig:controls_impact}} illustrate these simulated systems. This procedure allows us to study the effect on network functionality for varying control numbers and control positioning.\\
For each network, we sample $20,000$ control voltage combinations, producing $20,000$ sets of electric currents $(I_{00},I_{10},I_{01},I_{11})$ using KMC. Utilizing \mbox{Eq. (\ref{eq:fitness})}, we calculate the fitness of each control voltage combination for each Boolean logic gate at $\delta=0$. The fitness distributions for each gate across different networks are depicted as box plots in \mbox{FIG. \ref{fig:controls_impact} a)}, providing insights into the relationship between functionality and number of control electrodes.\\
First, in agreement with theoretical expectation, see  \mbox{Eq. (\ref{eq:f_expected}) and (\ref{eq:f_expected2})}, we observe similar distributions for the respective AND/OR, NAND/NOR, and XOR/XNOR pairs. Furthermore, the distributions of the first two pairs are just mirror images of each other. Any remaining variations in $<F>$ and $\text{Var}(F)$ within a pair can be attributed to covariance values between $M_i$ and $X$, which we did not consider in \mbox{Eq. (\ref{eq:f_expected})} and \mbox{Eq. (\ref{eq:f_expected2})}.\\
The impact on fitness when altering the number or position of control electrodes is significantly pronounced. We observe the most dramatic effect for controls in close proximity to the output electrode. The measures of nonlinear behavior, i.e., $Q_{\text{NDR}}$ and $Q_{\text{NLS}}$, exhibit much larger values, see \mbox{FIG. \ref{fig:controls_impact} b)}. Theoretically, one expects a strong correlation between $Q_{\text{NDR}}$ and the availability of high-fitness NAND/NOR gates as well as between $Q_{\text{NLS}}$ and the number of exclusive gates XOR/XNOR. Indeed, as seen from the box plots, a significant correlation is indeed observed. \\
As a secondary effect, it turns out to be helpful to use the maximum number of control electrodes $N_{C} = 9$. Only in  the case of AND/OR gates, a reduction in the number of controls generally leads to an increase in average fitness, while retaining electrodes near the output results in the most significant second moment.\\
Moreover, we observe a strong impact on the correlation coefficient between $M_l$ and $M_r$. The  presence of control electrodes adjacent to the output electrode as well as an increase in the number of control electrodes significantly reduces this correlation, going along with a stronger independence of $I_{01}$ and $I_{10}$.\\

As the results in \mbox{FIG. \ref{fig:controls_impact}} suggest a substantial impact of the positioning of controls relative to the output or inputs, we performed two additional simulations for a network of size $7 \times 7$.  First, we randomly adjusted the voltages of all 7 electrodes and determined the resulting correlation between individual voltages and the output current. The Pearson correlations are illustrated in \mbox{FIG. \ref{fig:electrode_sens}}. In alignment with previous observations, electrodes in close proximity to the output electrode (specifically $E_5$/$E_6$) exhibit a strong impact via notable correlations between electrode voltages and output electric current $I$. This observation is further supported by the current-voltage dependencies, which are strongest when altering electrode voltages near the output (see \mbox{FIG. \ref{fig:iv_curves})}.\\
Second, we selected inputs at all possible electrode combinations $(E_i, E_j)$ and sampled multiple control voltage combinations for any possible input electrode voltage combination $\{(E_i, E_j), (E_i+\Delta, E_j), (E_i, E_j+\Delta), (E_i+\Delta, E_j+\Delta)\}$ with $\Delta = 10 \text{ mV}$. The last two plots in \mbox{FIG. \ref{fig:electrode_sens}} depict the nonlinear features of NDR and NLS for each input position combination. If a single input is positioned adjacent to the output at $E_5$ or $E_6$, NLS is lost entirely due to the strong correlation with the electric current. For either NDR or NLS, optimal results are achieved by breaking symmetry and attaching one electrode at the corner opposite to the output electrode at position $E_0$, with the second input placed elsewhere. In fact, there is not a significant difference in attaching the second input at $E_1$, $E_2$, $E_3$, or $E_4$. The advantage of having an asymmetric electrode connection may indicate a potential functionality gain for disordered networks in general.

\begin{figure*}
\centering
\textbf{a)}
\includegraphics[width=0.3\textwidth]{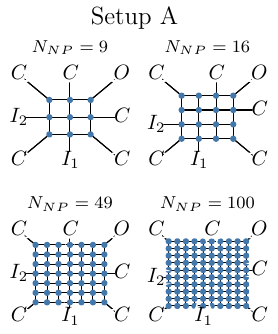}
\includegraphics[width=0.3\textwidth]{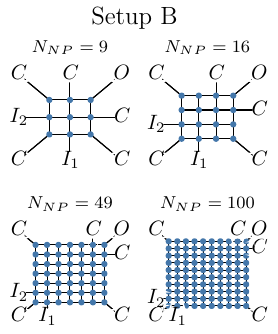}
\textbf{b)}
\includegraphics{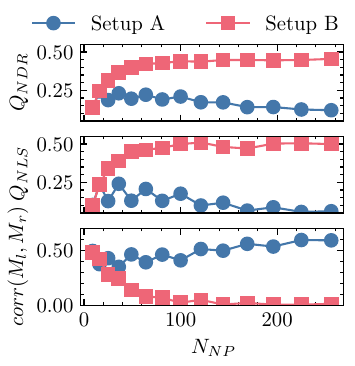}
\caption{\label{fig:ss_influence}
Simulation results for negative differential resistance ($Q_{\text{NDR}}$), nonlinear separability ($Q_{\text{NLS}}$), and input mobility correlation ($corr(M_l,M_r)$) in networks with varying numbers of NPs ($N_{\text{NP}}$). These networks are surrounded by five control electrodes ($C$), two input electrodes ($I_1$,$I_2$), and one output electrode ($O$). Two setups are considered: Setup A maintains same relative distances between all electrodes, while Setup B keeps the electrodes at fixed positions.}
\end{figure*}

\subsection{System Size}
\label{sec:system_size}

In this section, we investigate the influence of system size on the network functionality of regular grid-like NP networks. Following the experimental device proposed in \cite{bose2015}, each network is surrounded by a configuration of \mbox{8 electrodes}. We connect the output to one corner of the network, and both inputs to the opposite edges to ensure that most NPs are between the input and output. This leaves five control electrodes to be attached at the remaining corners or edges.\\
Considering the strong relation to electrode positioning indicated in the previous section, we distinguish between two setups, as displayed in \mbox{FIG. \ref{fig:ss_influence} a)}. In Setup A, we maintain consistent relative distances between all electrodes, connecting them either to a network corner or to the center of a network edge. In Setup B, the two electrodes, adjacent to the output, keep this minimum distance independent of system size.\\
Each simulation covers a specific system size, ranging from a $3 \times 3$ grid of NPs ($N_{\text{NP}} = 9$) to a $16 \times 16$ grid of NPs ($N_{\text{NP}} = 256$). We conduct $20,000$ simulations for each system size, sampling control voltage combinations and producing sets of electric currents $(I_{00}, I_{10}, I_{01}, I_{11})$. \\

In \mbox{FIG. \ref{fig:ss_influence} b)}, we break down the fitness into its key nonlinear components: $Q_{\text{NDR}}$, $Q_{\text{NLS}}$, and $corr(M_l,M_r)$. There is a noticeable difference between both setups. In Setup B, a saturation behavior is observed for each nonlinear component. For $N_{\text{NP}} > 100$, maximum NDR and NLS are achieved, and both inputs start to become fully independent. In fact, $Q_{\text{NDR}}$ even reaches its theoretical upper bound $Q_{\text{NDR}} = 0.5$ as described in \mbox{Section \ref{sec:ana_methods}}. In contrast, Setup A shows a less pronounced dependence. For systems seeking nonlinear separability, those with more than $50$ NPs are not preferred. There is a decline in performance for increasing numbers of NPs, suggesting that smaller systems tend to perform better in terms of nonlinear features, provided there is a minimum of about $16$ NPs. Regarding the correlation between both mobility values $M_l$ and $M_r$, we observe that independence between input electrodes is never achieved, with $corr(M_l,M_r) \approx 0.5$ for all networks. As shown in   \mbox{Supplementary Section \ref{sec:supp_delta}} and \mbox{FIG. \ref{fig:box_vs_size_setup}} the fitness distributions for AND, NAND, and XOR follow the same trend.  \\
Similar to the findings in the previous section, we observe that control electrodes in close proximity to the output electrode play a pivotal role in overall network functionalities, particularly in terms of nonlinearity. As the system size increases in Setup A, the distance between the output and its closest control electrodes also increases, leading to a negative impact on performance. Despite achieving better results in larger systems with Setup B, it is crucial to note that in practical experiments, the network configuration may not resemble Setup B. Therefore, we argue that in experiments, one should be cautious about making the network too large, even though theoretically, saturation in nonlinearity should occur. Optimal results are obtained when aiming for a maximum of about $N_{\text{NP}} = 100$, especially if electrodes can be densely packed. Otherwise, even a smaller system may prove to be more beneficial.\\
It is noteworthy that the saturation observed in \mbox{FIG. \ref{fig:ss_influence}} occurs only for $\delta = 0$, as $Q_{\text{NDR}}$ and $Q_{\text{NLS}}$ have been derived under this assumption (see \mbox{Section \ref{sec:ana_methods}}). For larger values of $\delta$, indicating a contribution of the logic gates offsets to the fitness value, we deviate from the saturation behavior, showing a clear maximum in fitness at about $N_{\text{NP}} = 100$. Considering the preference for logic gates without an offset in the experimental setup, this observation further supports the identification of an optimal system size at or below $100$ NPs.

\section{\label{sec:conclusion}Conclusion\protect}

We have developed a versatile and efficient kinetic Monte Carlo simulation tool designed to model charge tunneling dynamics in NP networks. Building upon established concepts of single electronics, the model has been specifically tailored for NP networks. The tool allows for the investigation of larger networks, comprising hundreds of NPs, within reasonable processing times.\\
Within our simulation framework, electrodes can be connected to the network at arbitrary positions and designated as input, output, or control electrodes. We have successfully mapped Boolean logic gate functionality to the dependence between input and output electrodes. Adjusting the control electrode voltages, the charge tunneling dynamics is manipulated, enabling the same network to be configured into any desired logic gate.\\
We assessed the quality of the network in functioning as a logic gate through a fitness function. Increasing the number of control electrodes consistently enhances fitness, with controls in close proximity to the output electrode proving most pivotal. When altering the positions of inputs and controls, our results demonstrated a loss of functionality when inputs are positioned in close proximity to the output. Additionally, we observed that an asymmetric connection of inputs could be advantageous.\\
We further related the fitness of each logic gate to a novel set of variables, enabling the derivation of measurement tools for key qualitative nonlinear features such as negative differential resistance and nonlinear separability. These features, while essential for Boolean logic, also hold relevance for other classification or machine learning functionalities.\\
Upon increasing the network's size, we unveil optimal nonlinear properties for a minimum of about $100$ nanoparticles. However, this saturation needs electrodes to remain close to the network's output. If an increase in size is accompanied by an enlargement of the area between the output and its neighboring electrodes, a clear maximum is achieved below $100$ nanoparticles. The same holds true when the logic gate offset is also considered in the fitness. In this context, we argue that smaller networks, $100$ nanoparticles at maximum, are advantageous in experimental setup where ensuring the correct electrode connection distance (i.e. nanoparticles in between) may be challenging.\\
The results in this work describe the impact of constant voltage signals on the network response. An important next step involves time-varying signals, aligning the input time scales with the intrinsic time scales of the network states. This should allow us to explore memory effects, interpreting the network as a dynamical system due to its recurrent connection topology. Subsequently, the potential application of reservoir computing allow us to utilize NP networks for temporal signal processing, encompassing tasks such as time series forecasting and approximating not only functions, as Boolean logic, but also dynamical systems.\\
Given the findings of this work on the influence of disorder and asymmetry, the study of disordered network topologies or variable NP sizes, resistances or materials may also provide new interesting facets of brain-inspired applications.

\section{\label{sec:acknowledgement}Acknowledgement\protect}

This work was funded by the Deutsche Forschungsgemeinschaft (DFG, German Research Foundation) through project 433682494--SFB 1459. Furthermore, we are grateful for helpful discussions with M. Beuel, P. A. Bobbert, M. Gnutzmann, B. J. Ravoo, L. Schlichter, H. Tertilt, and E. Wonisch about the topic of this paper.


\bibliographystyle{apsrev4-1}
\bibliography{paper_temp}


\onecolumngrid
\renewcommand{\thefigure}{S\arabic{figure}}
\setcounter{figure}{0}
\section{\label{sec:supp} Supplementary}

\subsection{Capacitance of two adjacent Nanoparticles}
\label{sec:supp_i_c}
\begin{figure*}
\centering
\includegraphics[width=0.8\textwidth]{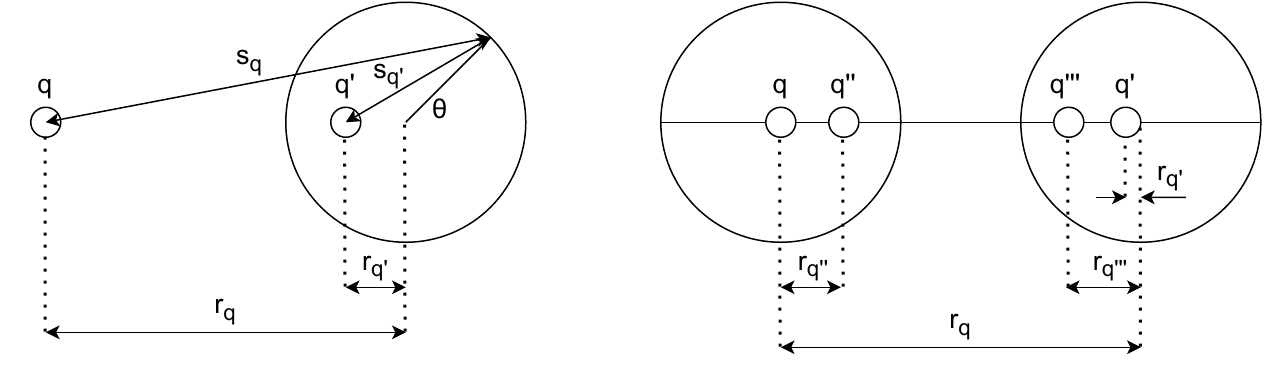}
\caption{\label{fig:capacitance}
Sketch illustrating the image charge method applied to a single grounded conducting sphere and two adjacent conducting spheres.}
\end{figure*}

First we want to find the potential $\phi(\textbf{r})$ inside a grounded conducting sphere (NP) with radius $R$ using the image charge method (see. \mbox{FIG. \ref{fig:capacitance}}). We introduce an image charge $q'$ for the charge $q$ positioned at $r_q$ relative to the origin of the sphere with distance $s_q$ to an arbitrary surface point on the sphere. The image charge is positioned along the line between the sphere center and charge $q$ at $r_{q'}$ and distance $s_{q'}$. We have to meet the boundary condition of zero potential on the spherical surface, i.e.

\begin{equation}
    V(s_{q},s_{q'}) \propto \frac{q}{s_{q}} + \frac{q'}{s_{q'}} \stackrel{!}{=} 0.
    \label{ic_0}
\end{equation}

For the two specific surface points aligned on the same axis as $q$ and $q'$, we get the vanishing potentials

\begin{eqnarray}
    \frac{q}{r_q + R} + \frac{q'}{r_{q'} + R} &=& 0 \nonumber\\ 
    \frac{q}{r_q - R} + \frac{q'}{R - r_{q'}} &=& 0.
    \label{ic_1}
\end{eqnarray}

for the following relationships:

\begin{eqnarray}
    q' &=& -\frac{R}{r_q} \cdot q \nonumber\\
    r_{q'} &=& \frac{R^2}{r_q}.
    \label{ic_2}
\end{eqnarray}

Now, for an arbitrary surface points on the sphere with distances to the first charge $s_q$ and second charge $s_{q'}$, we get vanishing potentials when inputting 

\[ s_q = \sqrt{r_q^2 + R^2 + 2r_qR\cos(\theta)} \text{, } s_{q'} = \sqrt{r_{q'}^2 + R^2 + 2r_{q'}R\cos(\theta)} \]

into \mbox{Eq. (\ref{ic_0})} with \mbox{Eq. (\ref{ic_2})}.\\
With these results, one can now calculate the total capacitance of two equally shaped spheres (see. \mbox{FIG. \ref{fig:capacitance}}). Again using the image charge method, both spheres will be replaced by charges while maintaining the surfaces as equipotentials. The charge $q$ makes the left sphere an equipotential, disturbing the right sphere’s potential. The charge $q'$ leads to a zero equipotential of the right sphere but disturbing the left. The image of the image $q''$ compensates for $q'$ and so on. This procedure leads to the series

\begin{equation}
    Q = q \cdot \left(1 + \frac{R^2}{r_q^2 - R^2} + \frac{R^4}{r_q^4 - 3r_q^2R^2 + R^4} + \dots \right)
    \label{ic_3}
\end{equation}

and with the potential $\phi = \frac{q}{4 \pi \epsilon R}$ to the total capacitance

\begin{equation}
    C_\Sigma = \frac{\partial Q}{\partial \phi} = 4 \pi \epsilon R \cdot \left(1 + \frac{R^2}{r_q^2 - R^2} + \frac{R^4}{r_q^4 - 3r_q^2R^2 + R^4} + \dots \right).
    \label{ic_4}
\end{equation}

Finally, separating the total capacitance in mutual capacitance $C_{ij}$ between NPs $i$ and $j$ and self-capacitance $C_{i,\text{self}}$ of NP $i$, and substituting the distance between both spheres into \mbox{Eq. (\ref{ic_4})}, the final equations

\begin{equation}
    C_{ij} = 4 \pi \epsilon_0 \epsilon_{\text{m}} \frac{r^2}{2r + d} \left(1 + \frac{r^2}{(2r + d)^2 - 2r^2} + \frac{r^4}{(2r + d)^4 - 4(2r+d)^2r^2 + 3r^4} + \dots \right).
    \label{ic_5}
\end{equation}

\begin{equation}
    C_{i,\text{self}} = 4 \pi \epsilon_0 \epsilon_{\text{SiO}_{2}} r \left(1 - \frac{r}{2r + d} + \frac{r^2}{(2r + d)^2 - 2r^2} + \frac{r^3}{(2r + d)^3 - 2(2r+d)r^2} + \dots \right).
    \label{ic_6}
\end{equation}

are achieved, with $r_q \rightarrow 2r + d$ as NP radius $r$ and spacing in between NPs $d$.\\

\subsection{Current voltage dependence}
\label{sec:supp_iv}

\begin{figure*}
\centering
\includegraphics[width=\textwidth]{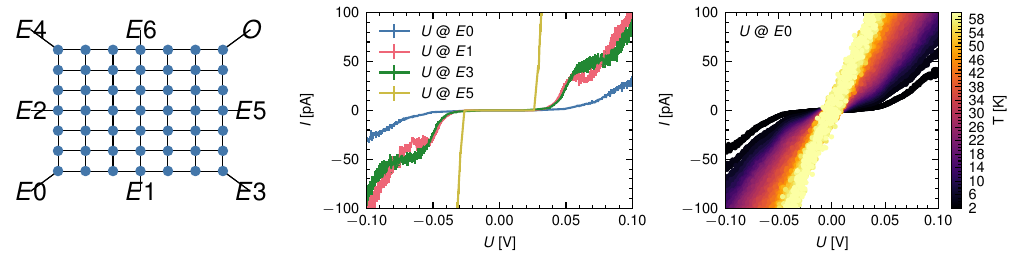}
\caption{\label{fig:iv_curves}
Current-voltage (I-V) dependence for a network of $49$ nanoparticles with $8$ electrodes. A voltage is applied to the input electrode, while all other electrodes are grounded. The left figure shows I-V curves with the input electrode positioned at various locations. The right figure shows I-V curves for a fixed input position under different operating temperatures ($T$).}
\end{figure*}

For a fixed $N_{\text{NP}} = 49$ network with $8$ electrodes, we measured the output electrode electric current ($I$) with respect to the input electrode voltage ($U$). The output electrode, along with all other electrodes except the input, is grounded. The middle plot in \mbox{FIG. \ref{fig:iv_curves}} shows current-voltage dependencies obtained by varying the position of the input electrode. We observe a significant increase in current after surpassing the blockade regime when attaching the input close to the output at $E_5$. Accordingly, repositioning the input farther away at $E_1$ or $E_2$ results in a substantial decreases in current, indicating that most of the electric current might be drained at $E_5$. For input at $E_0$, the slope is the shallowest.\\
The right plot in \mbox{FIG. \ref{fig:iv_curves}} maintains the input at $E_0$ and uses different colors to represent various network temperatures. Here, it becomes evident that increasing the temperature above approximately \mbox{$20$ K} leads to a linearization of the current-voltage dependence, indicating the loss of functionality due to the loss of the network's nonlinear activation functions. Both results are in well agreement with the measurement results from \cite{bose2015}.

\subsection{Voltage re-scaling based on System Size}
\label{sec:supp_scaling}

\begin{figure*}
\centering
\includegraphics[width=\textwidth]{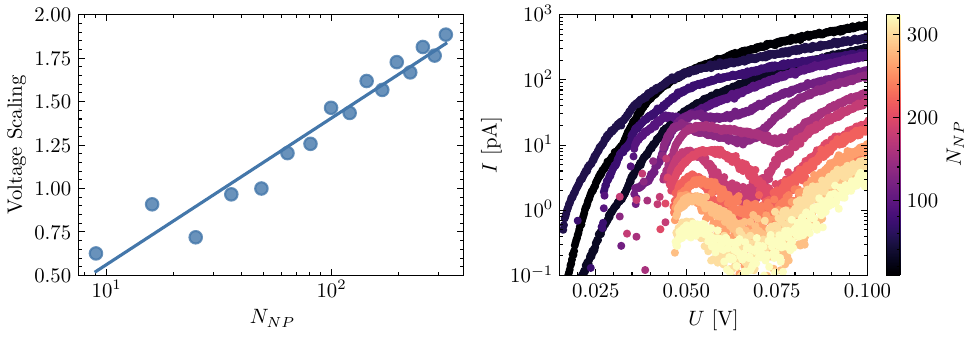}
\caption{\label{fig:volt_scaling}
Voltage scaling for different numbers of NPs ($N_{\text{NP}}$). The factor determines how the control sampling range and input states are adjusted depending on the network size. The factor results from the current-voltage (I-V) dependence. When a voltage ($U$) is multiplied by the factor each system produces approximately the same electric current ($I$).}
\end{figure*}

As there is a voltage drop across the entire network, we have to scale the magnitude of electrode voltages according to the number of NPs. Without such scaling, voltage ranges that are above the Coulomb blockade regime for a small network may not correspondingly surpass blockade for a larger network. This would result in decreasing electric currents with increasing numbers of NPs, not allowing to assess the true impact of system size but merely emphasizing the impact of vanishing network potentials.\\
To address this, we conducted electric current measurements ($I$) of the output electrode for networks with $8$ electrodes. 
In all network setups, two input electrodes were positioned at opposite edges of the output electrode, both set to the same voltage ($U$), while all other electrodes remained grounded. The right plot in \mbox{FIG. \ref{fig:volt_scaling}} illustrates the $I$ to $U$ dependence for various numbers of NPs ($N_{\text{NP}}$). Simultaneously, the left plot presents the factor by which $U$ should be multiplied to maintain electric currents of the same magnitude. The current values were normalized to those of the $7 \times 7$ network. For all simulations, we then applied the $N_{\text{NP}}$-specific factor to the sampling ranges of either $U_I \in \{0.0, 10.0\}$ mV or $U_C \in [-50.0, 50.0]$ mV. In \mbox{FIG. \ref{fig:volt_scaling}} we also clearly see the phenomenon of negative differential resistance, as increasing the voltage reduces the output electric current, at least for a network of about $100$ nanoparticles at minimum.

\subsection{System Size and $\delta$-dependence}
\label{sec:supp_delta}

\begin{figure*}
\centering
\includegraphics[width=\textwidth]{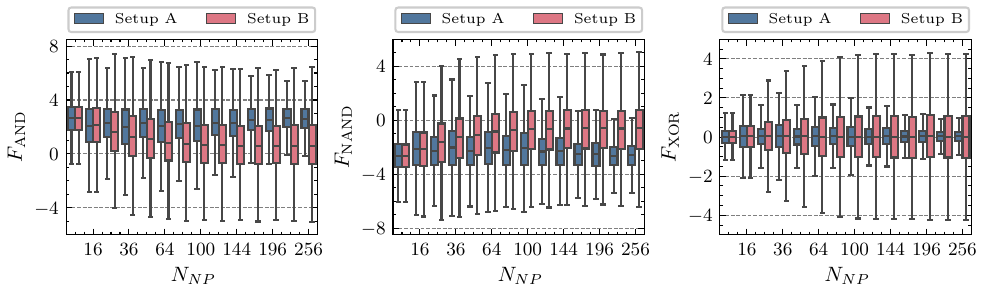}
\caption{\label{fig:box_vs_size_setup}
Box plots of fitness values for AND, NAND, XOR ($F_{\text{AND}}$, $F_{\text{NAND}}$, $F_{\text{XOR}}$) logic gates for systems with varying numbers of nanoparticles ($N_{\text{NP}}$). In each plot Setups A and B are compared. The length of a bar is equal to two standard deviations.}
\end{figure*}

\begin{figure*}
\centering
\includegraphics[width=\textwidth]{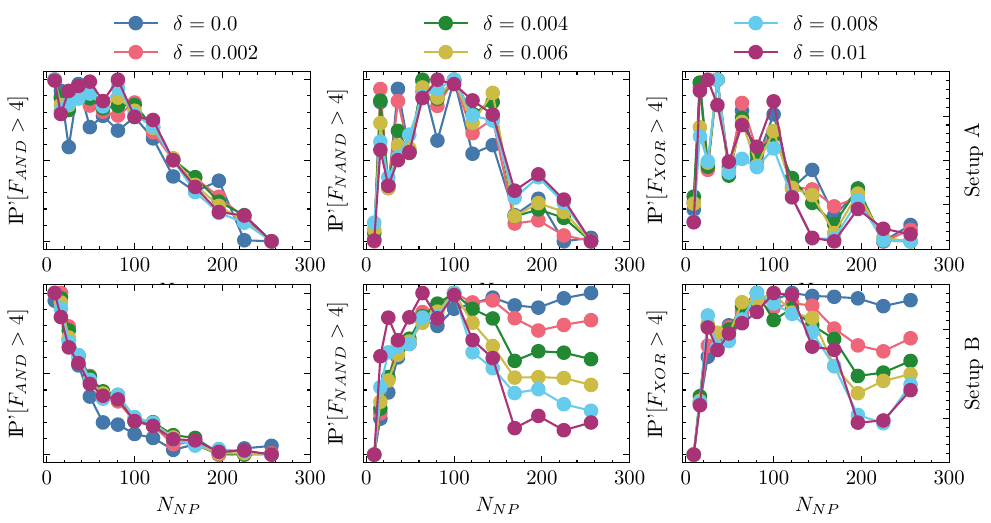}
\caption{\label{fig:delta_dep}
Influence of offset on fitness ($F$). For networks with varying numbers of nanoparticles ($N_{\text{NP}}$) the probability of finding an AND, NAND, or XOR logic gate with $F > 4$ is shown. Each probability is assessed while systematically varying the contribution of the offset to fitness by increasing the parameter $\delta$. The probabilities are normalized such that the maximum value for the different $\delta$-values is identical for every plot.}
\end{figure*}

Here, we present additional insights into the dependence of fitness ($F$) on system size. Box plots in \mbox{FIG. \ref{fig:box_vs_size_setup}} illustrate fitness values for AND, NAND, and XOR logic gates across networks with varying numbers of NPs ($N_{\text{NP}}$). We compare the results for both setups, A and B, as depicted in \mbox{FIG. \ref{fig:ss_influence} a)}. Consistently, we observe similar behavior on average between $F_{\text{NAND}}$ and $Q_{\text{NDR}}$, as well as between $F_{\text{XOR}}$ and $Q_{\text{NLS}}$. The behavior of $F_{\text{AND}}$ is complementary to that of $F_{\text{NAND}}$.\\
We further investigate the influence of the offset $|c|$ in the fitness definition:

\[ F = \frac{m}{\sqrt{MSE} + \delta \cdot |c|} \quad \text{(from Eq. (\ref{eq:fitness}))}. \]

As explained in \mbox{Section \ref{sec:ana_methods}}, the offset is determined by the average off-currents. While neglecting the offset in fitness evaluation is advantageous for theoretical purposes (as discussed in \mbox{Section \ref{sec:ana_methods}}), we also examine its impact on the occurrence of logic gate functionality, considering that having logic gates close to zero is beneficial in experimental setups. To explore this, we vary the factor $\delta$ in \mbox{Eq. (\ref{eq:fitness})} to modulate the weight of the $|c|$ contribution. In previous work \cite{bose2015}, an empirical value of $\delta = 0.01$ was used. \mbox{FIG. \ref{fig:delta_dep}} illustrates the probability $\mathbb{P}$ of finding AND, NAND, or XOR logic gates with $F > 4$ in networks with variable numbers of NPs ($N_{\text{NP}}$). For each logic gate, we compare results for both setups shown in \mbox{FIG. \ref{fig:ss_influence} a)}. Considering the trivial effect of increasing the denominator with rising $\delta$, we normalize the probability $\mathbb{P} \rightarrow \mathbb{P}'$ across all $\delta$ values to a range of $[0,1]$. This normalization facilitates a qualitative assessment of changes relative to $N_{\text{NP}}$. In Setup B, a noticeable influence is evident. We observe a loss of the saturation behavior for NAND and XOR logic gates with an increasing impact of $|c|$. This observation reinforces our contention that the optimal network size is around $100$ NPs, as evidenced by the maximum fitness achieved at this point. Logic gates in for Setup A remain unaffected.

\begin{figure*}
\centering
\includegraphics[width=\textwidth]{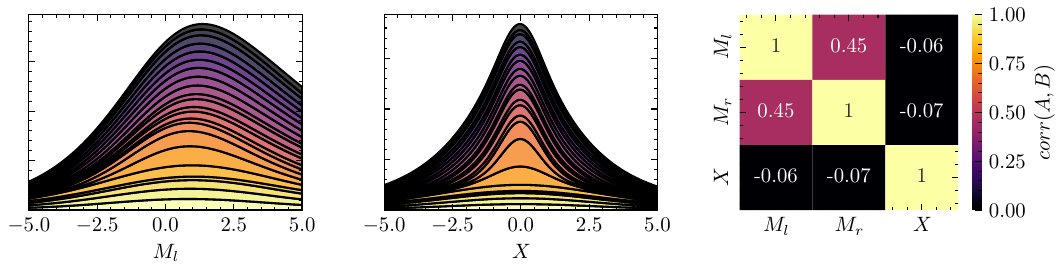}
\caption{\label{fig:x_and_r_dist}
Distributions and covariance heat map of $M_l$/$M_r$ and $X$ across arbitrary systems with varying numbers of nanoparticles and control electrodes.}
\end{figure*}

\begin{figure*}
\centering
\includegraphics[width=\textwidth]{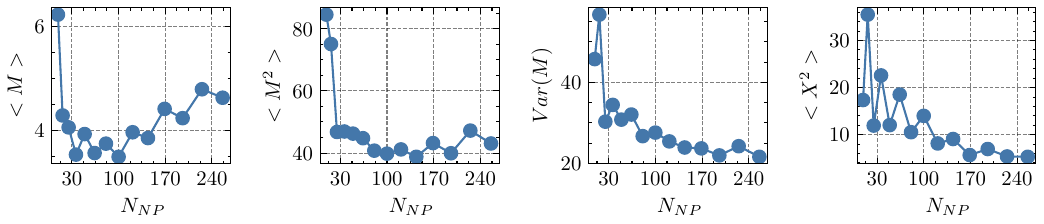}
\includegraphics[width=\textwidth]{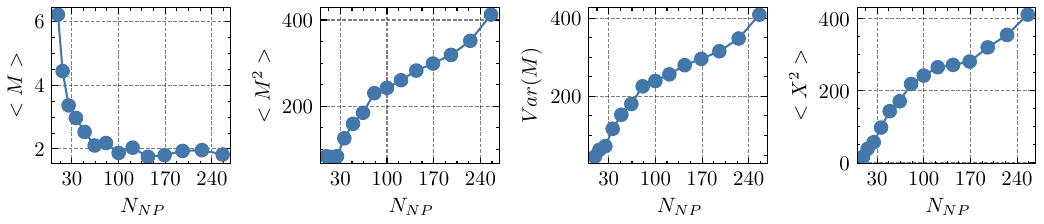}
\includegraphics[width=\textwidth]{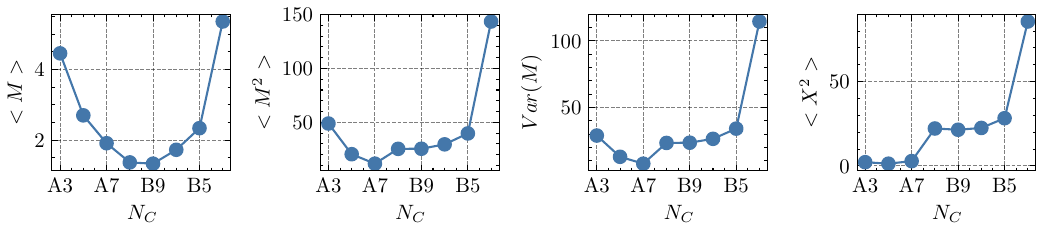}
\caption{\label{fig:fitness_parameter}
Impact of number of nanoparticles ($N_{\text{NP}}$) and number of control electrodes ($N_{C}$) on expected values of all variables derived from the fitness decomposition. The first row displays the $N_{\text{NP}}$ results for \mbox{Setup A}, while the second row covers \mbox{Setup B}.}
\end{figure*}

\subsection{Parameter behavior}
\label{sec:supp_paras}

For NP networks of varying system sizes and numbers of control electrodes, we display the distribution of either $M_l$ or $M_r$ in the left plot of \mbox{FIG. \ref{fig:x_and_r_dist}}. We detect non-zero means with variable second or even third moment. Additionally, for the same set of simulations, we display the distribution of $X$ in the middle plot of \mbox{FIG. \ref{fig:x_and_r_dist}}, showing an approximately zero mean and variable second moment. The right plot in \mbox{FIG. \ref{fig:x_and_r_dist}} color-codes the Pearson correlation of these observables, revealing negligible correlations between $M_l$/$M_r$ and $X$. Based on these results we assume $<X> = 0$, $<M_l> = <M_l>$, $<M_l^2> = <M_l^2>$ and $\text{cov}(M_l,X) = \text{cov}(M_r,X) = 0$ for our calculations.\\
In the first row of plots in \mbox{FIG. \ref{fig:fitness_parameter}} we assess all expected values corresponding to $<F>$ and $<F^2>$ for variable numbers of NPs ($N_{\text{NP}}$) for Setup A in \mbox{FIG. \ref{fig:ss_influence} a)}. In the second row, we repeat this analysis for Setup B in \mbox{FIG. \ref{fig:ss_influence} a)}, while the last row corresponds to variable numbers of control electrodes $N_{C}$ according to the network sketches in \mbox{FIG. \ref{fig:controls_impact}}.

\end{document}